\documentclass[%
 reprint,
superscriptaddress,
 amsmath,amssymb,
 aps,
]{revtex4-2}

\usepackage{graphicx}
\usepackage{dcolumn}
\usepackage{bm}
\usepackage{chemformula}
\usepackage{amsmath,amsthm,amssymb}
\usepackage{gensymb}
\usepackage{xr}
\usepackage{hyperref}
\usepackage{soul}
\usepackage[mathlines]{lineno}

\begin{document}

\title{Mott resistive switching initiated by topological defects}

\author{Alessandra Milloch}
\email[]{alessandra.milloch@unicatt.it}
\affiliation{Department of Mathematics and Physics, Università Cattolica del Sacro Cuore, Brescia I-25133, Italy}
\affiliation{ILAMP (Interdisciplinary Laboratories for Advanced
Materials Physics), Università Cattolica del Sacro Cuore, Brescia I-25133, Italy}
\affiliation{Department of Physics and Astronomy, KU Leuven, B-3001 Leuven, Belgium}

\author{Ignacio Figueruelo-Campanero}
\email[]{ignacio.figueruelo@imdea.org}
\affiliation{IMDEA Nanociencia, Cantoblanco, 28049 Madrid, Spain}
\affiliation{Facultad Ciencias Físicas, Universidad Complutense, 28040 Madrid, Spain}

\author{Wei-Fan Hsu}
\affiliation{Department of Physics and Astronomy, KU Leuven, B-3001 Leuven, Belgium}

\author{Selene Mor}
\affiliation{Department of Mathematics and Physics, Università Cattolica del Sacro Cuore, Brescia I-25133, Italy}
\affiliation{ILAMP (Interdisciplinary Laboratories for Advanced
Materials Physics), Università Cattolica del Sacro Cuore, Brescia I-25133, Italy}

\author{Simon Mellaerts}
\affiliation{Department of Physics and Astronomy, KU Leuven, B-3001 Leuven, Belgium}

\author{Francesco Maccherozzi}
\affiliation{Diamond Light Source, Didcot, Oxfordshire OX11 0DE, UK}

\author{Larissa {Ishibe Veiga}}
\affiliation{Diamond Light Source, Didcot, Oxfordshire OX11 0DE, UK}

\author{Sarnjeet S. Dhesi}
\affiliation{Diamond Light Source, Didcot, Oxfordshire OX11 0DE, UK}

\author{Mauro Spera}
\affiliation{Department of Mathematics and Physics, Università Cattolica del Sacro Cuore, Brescia I-25133, Italy}

\author{Jin Won Seo}
\affiliation{Department of Materials Engineering, KU Leuven, 3001 Leuven, Belgium}

\author{Jean-Pierre Locquet}
\affiliation{Department of Physics and Astronomy, KU Leuven, B-3001 Leuven, Belgium}

\author{Michele Fabrizio}
\affiliation{Scuola Internazionale Superiore di Studi Avanzati (SISSA), Via Bonomea 265, 34136 Trieste, Italy}

\author{Mariela Menghini}
\affiliation{IMDEA Nanociencia, Cantoblanco, 28049 Madrid, Spain}

\author{Claudio Giannetti}
\email[]{claudio.giannetti@unicatt.it}
\affiliation{Department of Mathematics and Physics, Università Cattolica del Sacro Cuore, Brescia I-25133, Italy}
\affiliation{ILAMP (Interdisciplinary Laboratories for Advanced
Materials Physics), Università Cattolica del Sacro Cuore, Brescia I-25133, Italy}
\affiliation{CNR-INO (National Institute of Optics), via Branze 45, 25123 Brescia, Italy}

\begin{abstract}
\textbf{Resistive switching is the fundamental process that triggers the sudden change of the electrical properties in solid-state devices under the action of intense electric fields \cite{Wang2020}. Despite its relevance for information processing, ultrafast electronics, neuromorphic devices, resistive memories and brain-inspired computation \cite{Zhou2015,Tokura2017,delValle2018,Adda2018,delValle2019,Perez2019,Wang2020,DelValle2020,Zhang2021,Deng2021,Deng2023,Ran2023,yang2011oxide,Mehonic2022}, the nature of the local stochastic fluctuations that drive the formation of metallic nuclei out of the insulating state has remained hidden.}

\textbf{Here, using $operando$ X-ray nano-imaging, we have captured the early-stages of resistive switching in a V$_2$O$_3$-based device under working conditions. V$_2$O$_3$ is a paradigmatic Mott material \cite{Tokura2017}, which undergoes a first-order metal-to-insulator transition coupled to a lattice transformation that breaks the threefold rotational symmetry of the rhombohedral metal phase \cite{Zhou2015,Adda2018,delValle2019,DelValle2020,Zhang2021,Deng2021,Deng2023,hsu2023raman}. 
We reveal a new class of volatile electronic switching triggered by nanoscale topological defects of the lattice order parameter of the insulating phase. 
Our results pave the way to the use of strain engineering approaches to manipulate topological defects and achieve the full control of the electronic Mott switching. The concept of topology-driven reversible electronic transition is of interest for a broad class of quantum materials, comprising transition metal oxides, chalcogenides and kagome metals, that exhibit first-order electronic transitions coupled to a symmetry-breaking order.}
\end{abstract}

\maketitle

\begin{figure*}[]
\includegraphics[width = 17 cm]{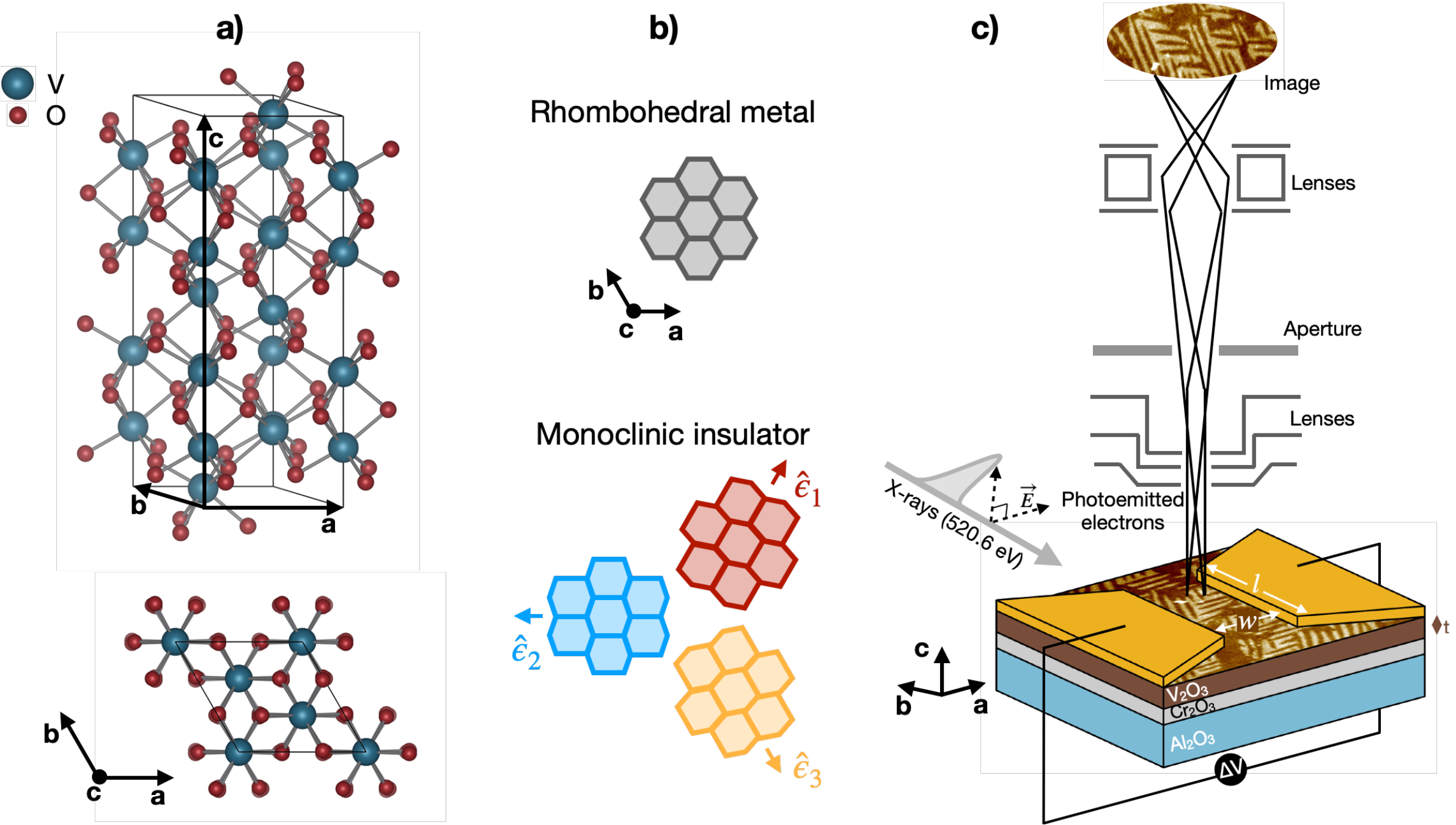}
\caption{a) Non-primitive hexagonal unit cell of the high-temperature rhombohedral phase. b) Sketch of the rhombohedral-to-monoclinic distortion taking place along each of the three equivalent hexagonal axes. c) PEEM experimental setup. X-ray radiation with tunable energy resonant with the vanadium L$_{2,3}$ edge impinges on the sample and the emitted electrons are collected and imaged through electrostatic and magnetic lenses. The \ch{V_2O_3} film of thickness $t$ = 20 nm is coated with gold metal electrodes separated by a 2 \textmu m width ($w$) $\times$ 50 \textmu m length ($l$) gap, allowing to drive a current through the device while simultaneously acquiring XLD-PEEM images.}
\label{fig: setup}
\end{figure*}

The insulator-to-metal transition (IMT) in Mott materials is a key mechanism for the development of the next generation Mottronic devices \cite{Tokura2017,yang2011oxide}. The intrinsic correlated nature of the Mott insulating state makes these systems fragile to external stimuli \cite{zhang2014dynamics,basov2017towards}, such as the application of an electric field, which can drive the collapse of the electronic band structure and the sudden release of a large number of free carriers \cite{mazza2016field,Suen2023}. At the macroscopic level, this phenomenon manifests in the resistive switching process, i.e., a sharp increase of the current flow when the applied voltage exceeds 
a threshold value \cite{Wouters2019,Kalcheim2020,stoliar2013,delValle2019,guiot2013avalanche,nakamura2013electric,fursina2009origin,fursina2012statistical,janod2015resistive}. This strong non-linearity triggered many efforts to develop neuromorphic building blocks for the hardware implementation of neural networks \cite{Mehonic2022} or for ultrafast volatile and non-volatile memories or processors \cite{waser2007nanoionics,ielmini2016resistive,Ran2023}.    
The state-of-the-art macroscopic models \cite{Lee2015} are based on resistor networks that consider interconnected nodes transforming from the insulating to metallic state in the presence of an electric field. Above a certain threshold, a percolative transition takes place, thus leading to the formation of a conductive filament and a sudden resistivity drop \cite{stoliar2013,Stoliar2017}. 

The full control and, therefore, the exploitation of this process is prevented by the limited knowledge of the early-stage firing dynamics. Microscopically, little is known about the nature of the nanoscale regions that trigger the avalanche process. Also the relation between the electronic and structural properties of the switched regions and those of the pristine insulating template is a matter of debate. Pioneering optical microscopy experiments captured the real-time formation of macroscopic metallic channels \cite{delValle2018,lange2021imaging,del2021spatiotemporal,Babich2021,Luibrand2023}, while lacking the resolution and sensitivity to address the origin of the switching process. 

Here, we introduce a resonant X-ray microscopy to take nanoscale snapshots of the switching dynamics at the nanoscale during the application of an electric field to a V$_2$O$_3$-based nanodevice. Our results unveil the fundamental role played by the topology of the underlying lattice nanotexture. The breaking of the $C_3$ symmetry upon transition to the insulating monoclinic phase leads to the formation of three twin domains whose boundaries are oriented along the three hexagonal directions \cite{ronchi2019early,ronchi2022nanoscale}. The geometrical constraints produce strain topological defects at the corners of monoclinic domains crossing with an angle of 60$^{\circ}$. These nanoscale topological defects act as seeds of the metallic phase, thus triggering the macroscopic volatile switching. The nature of the transient metallic state is discussed  in view of achieving ultrafast and reversible all-electronic switching.

\ch{V_2O_3} is a prototypical Mott insulator that undergoes a thermally-driven transition from a high-temperature paramagnetic metal with rhombohedral lattice symmetry to a low-temperature antiferromagnetic insulator with monoclinic structure \cite{mcwhan1969mott,mcwhan1970metal,mcwhan1973metal}. 
The lattice transformation across the critical temperature $T_{IMT}$ implies the breaking of the $C_3$ symmetry of the non-primitive hexagonal unit cell of the high-temperature phase (see Fig. \ref{fig: setup}a). The structural transition can thus be described \cite{ronchi2022nanoscale} by a vector order parameter:
\begin{equation}
 \label{order_parameter}
 \overrightarrow{\epsilon}=(\epsilon_{31},\epsilon_{23})=\epsilon\,\big(\mathrm{cos}\phi_n, \mathrm{sin}\phi_n
 \big)
\end{equation}
associated to the shear strain components $\epsilon_{31}$ and $\epsilon_{23}$ that characterize the monoclinic distortion. Below $T_{IMT}$, the amplitude of the order parameter, $\epsilon$, becomes non-zero, while the phase can assume three different values:
\begin{equation}
 \label{phase}
\phi_n=(2n-1)\frac{\pi}{3}
\end{equation}
corresponding to the distortion along the three equivalent hexagonal axes of the rhombohedral phase, indicated in the following by the versors $\hat{\epsilon}_n$, $n$=1,2,3 (see Fig. \ref{fig: setup}b).

Resistive switching from the insulating to the metallic state can be induced by applying an electric field across patterned micro-gaps at temperatures close to $T_{IMT}$ \cite{guenon2013electrical,delValle2018,del2021spatiotemporal}. The resistive switching device investigated here is formed by a 20 nm \ch{V_2O_3} film coated with gold electrodes. \ch{V_2O_3} is grown by oxygen-assisted Molecular Beam Epitaxy on a (0001)-\ch{Al_2O_3} substrate with a 40 nm \ch{Cr_2O_3} buffer layer to reduce any interfacial residual strain \cite{Dillemans2014}. The resulting \ch{V_2O_3} film has the $c$ axis oriented parallel to the surface normal and undergoes the IMT at $T_{IMT}$ = 145 K (see Supplementary Information Fig. S1). Two gold electrodes allow the application of an electric bias across the gap of width $w$ = 2 \textmu m and length $l$ = 30 \textmu m (figure \ref{fig: setup}c). The gap region between the electrodes is imaged using Photo Emission Electron Microscopy (PEEM), combined with X-ray Linear Dichroism (XLD) at the \ch{L_{2,3}} vanadium edge (513-530 eV, see Figure S2) \cite{park2000spin,ronchi2019early,ronchi2022nanoscale}. The XLD-PEEM images are obtained from the normalized difference between images recorded with the light electric field vector, $\overrightarrow{E}$, parallel and perpendicular to the surface normal at a photon energy 520.6 eV. Since the XLD signal depends on the angle between the in-plane component of $\overrightarrow{E}$ and $\overrightarrow{\epsilon}(\mathbf{r})$ \cite{ronchi2022nanoscale} (see Fig. \ref{fig: setup}b and c), this technique provides a map - with $\sim$30 nm spatial resolution - of the three different monoclinic domains and their melting during the resistive switching process.

\begin{figure}[]
\includegraphics[width = 8.5 cm]{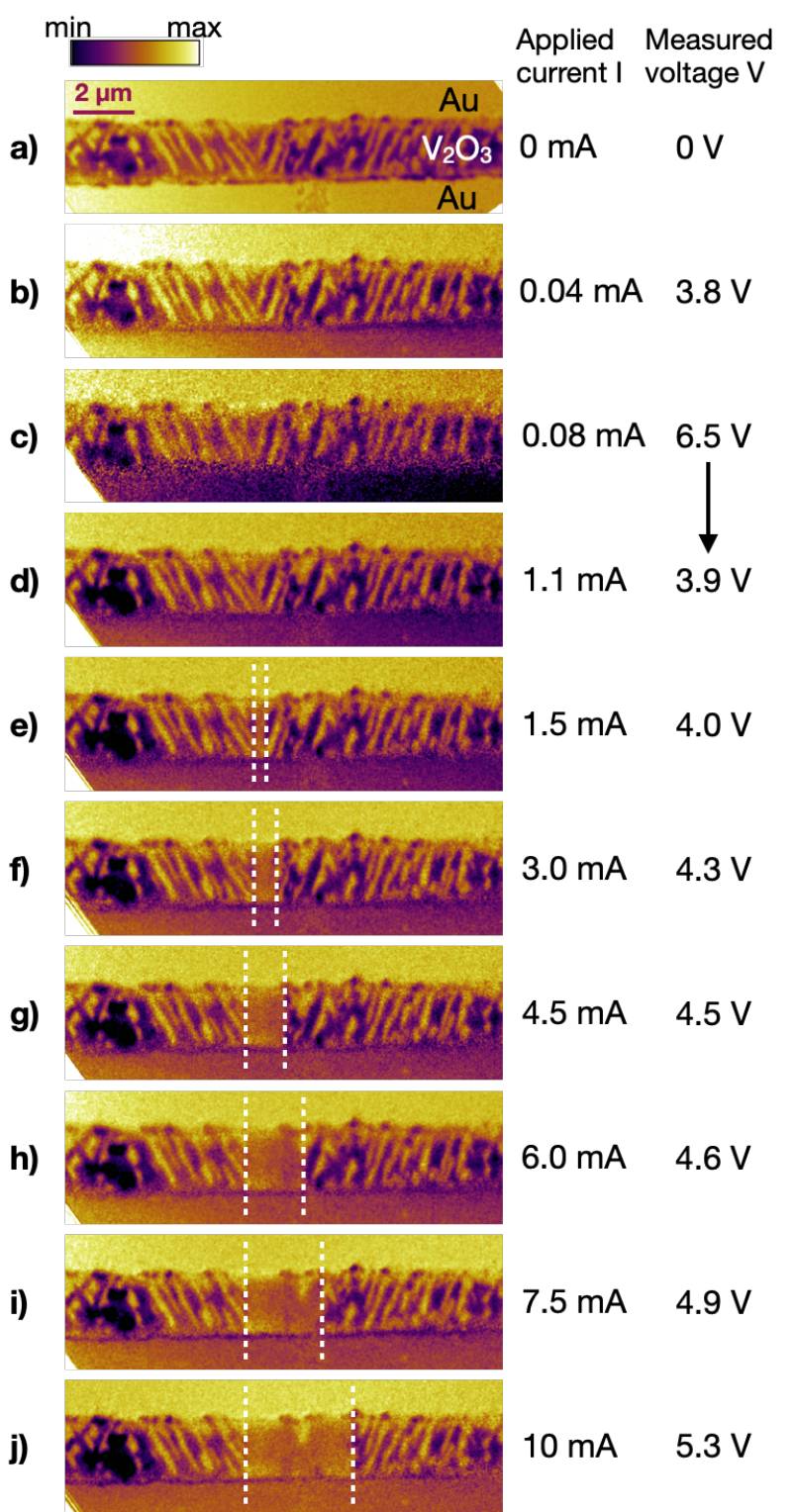}
\caption{PEEM images before (a) and during (b-j) electric current application at $T$ = 120 K. The homogeneous regions on top and bottom of each image are the gold electrodes; the area in between the electrodes represents the exposed \ch{V_2O_3} surface exhibiting the striped domains nanotexture typical of the antiferromagnetic monoclinic phase. For currents $I$ larger than 1.5 mA, the striped domains disappear in the region delimited by white dashed lines. This shows that a rhombohedral metallic filament is formed and widens as the current running through the device is increased.}
\label{fig: peem}
\end{figure}

Figure \ref{fig: peem}a) shows the PEEM image obtained in the monoclinic insulating phase of \ch{V_2O_3} at $T$ = 120 K. The in-gap \ch{V_2O_3} features the nanotexture typical of the monoclinic insulating phase \cite{ronchi2019early,ronchi2022nanoscale}. Monoclinic domains with different $\phi_n$ give rise to different XLD contrast, which can be appreciated as different color intensities within the PEEM image. The minimization of the total strain leads to the formation of stripe-like domains, whose directions are constrained by the symmetry of the system \cite{ronchi2022nanoscale}, as it will be discussed later. Each monoclinic insulating domain with a specific $\phi_n$, extends over a few micrometers in length, thus connecting the two electrodes, and it is characterized by a  lateral size of $w_{dom}\sim$200 nm \cite{ronchi2022nanoscale}.

The PEEM imaging is then repeated while driving a current $I$ through the device and measuring the voltage drop $V$ across the contacts. Figures \ref{fig: peem}b)-j) show the PEEM images acquired at increasing values of $I$, following the upward branch of the hysteresis cycle. The presence of an in-plane electric field across the electrodes introduces a weak image blurring that becomes significant for electric potential differences $V$ larger than 6-8 V. Despite this effect, the nanodomains are clearly resolved during the resistive switching process, which manifests itself in the voltage drop observed between 0.08 mA and 1.1 mA (Fig. \ref{fig: peem} c) and d) respectively). As the current is further increased beyond the threshold value necessary for the switching, we progressively observe the melting of the monoclinic nanotexture in the region delimited by white dashed lines (Fig. \ref{fig: peem} e)-j)) and the appearance of a channel with homogenous intensity. The XLD contrast measured in this region corresponds to the signal of the high-temperature rhombohedral phase. This is also confirmed from the angle dependence of the XLD signal \cite{ronchi2022nanoscale}. As shown in the Supplementary Information Fig. S3, images collected with two different angles of the X-rays polarization with respect to the in-plane \ch{V_2O_3} axes show no intensity variation upon sample rotation in the metallic channel, as opposed to the lateral monoclinic domains, whose signal depends on the angle between the light polarization and $ \overrightarrow{\epsilon}(\mathbf{r})$.
We can thus attribute the flat signal region in the middle of the gap to the formation of a metallic channel with rhombohedral lattice structure ($\epsilon$=0). In the Supplementary Information Fig. S4, we report PEEM images obtained when repeating the experiment in the same conditions but with a larger field of view that allows to capture the whole gap of the device. The metallic channel always forms in the same location within the gap and no other metallic paths are observed. Furthermore, when the applied current is removed, the metallic channel disappears and the monoclinic domains form again with the same pre-switching configuration, indicating a volatile process.  

\begin{figure}[h]
\includegraphics[width = 8 cm]{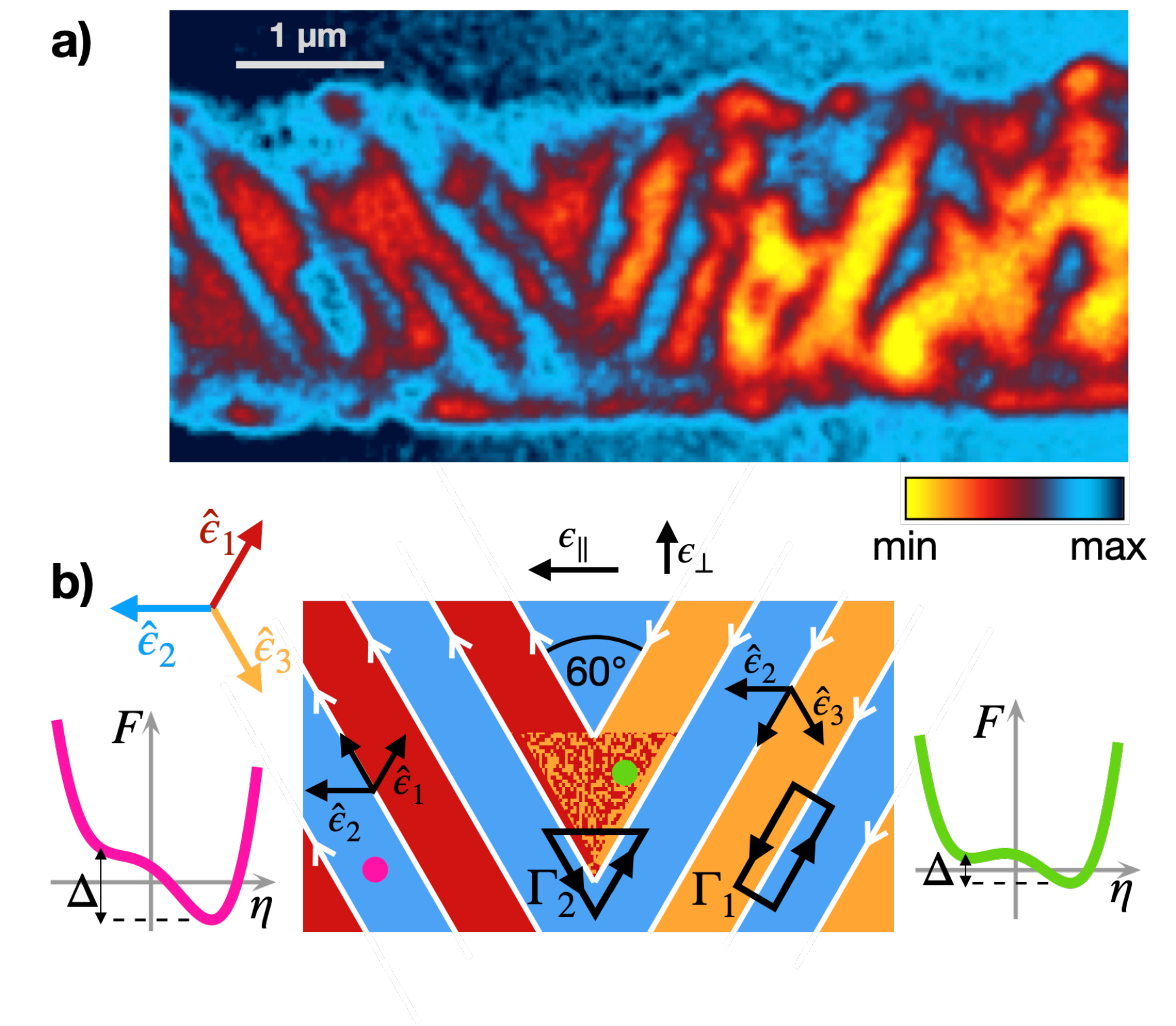}
\caption{a) Detail of the sample nanotexture in the region where the metallic filament is formed upon application of an above-threshold current. b) Sketch of the arrangement of monoclinic domains crossing at $60^\circ$ angle and forming a topological defect. Blue, red and yellow areas identify the three possible monoclinic domains corresponding to the three equivalent order parameter directions $\hat{\epsilon}_n$. The order parameter at the boundaries between different domains is oriented along $\hat{\epsilon}_1+\hat{\epsilon}_2$ ($2\pi/3$) for the red-blue interface and along $\hat{\epsilon}_2+\hat{\epsilon}_3$ ($4\pi/3$) for the blue-yellow interface. The mixed red-yellow triangular region indicates the local suppression of the strain at the topological defect. The energy functional $F$ as function of the order parameter $\eta$, sketched in the left and right panels, shows how a smaller value of $\epsilon^2$ due to the topological defect (green plot, solid line) results in a decrease of the insulator-metal energy difference $\Delta$.}
\label{fig: topology}
\end{figure}

The formation of the metallic channel is pinned by a specific topology of the lattice nanotexture, characterized by V-shaped domains, i.e., at the crossing point of domains with the same $\phi_n$ and directions that differ by a $\pi/3$ angle. In Fig. \ref{fig: topology}a) we report a detail of the switching region using a colorscale that highlights the nature of the three different domains: 1) light blue, which corresponds to a monoclinic distortion along the  $\hat{\epsilon}_2$ direction ($\phi_2$=$\pi$); 2) red, which corresponds to a monoclinic distortion along the  $\hat{\epsilon}_1$ direction ($\phi_1$=$\pi$/3); 3) yellow, which corresponds to a monoclinic distortion along the  $\hat{\epsilon}_3$ direction ($\phi_3$=5$\pi$/3).
As demonstrated in Ref. \citenum{ronchi2022nanoscale}, the stabilization of the monoclinic nanotexture is driven by the
Saint-Venant compatibility condition, which guarantees the continuity of the medium during a deformation and corresponds to the curl-free condition:
\begin{equation}
\label{SVcondition}
\overrightarrow{\nabla}\times\overrightarrow{\epsilon}(\mathbf{r})=0.
\end{equation}
The conservation of the parallel component of the position dependent order parameter, $\overrightarrow{\epsilon} (\mathbf{r})$, across an interface between two different domains, as implied by the curl-free condition, has two major consequences: 
\begin{itemize}
\item[i)] the interface between two different monoclinic domains must be oriented along the direction corresponding to the order parameter of the third domain; 
\item[ii)] the interface between a monoclinic and a rhombohedral metallic domain must be oriented perpendicularly to the order parameter of the monoclinic domain. 
\end{itemize}
If we consider, for example, a domain with order parameter along $\hat{\epsilon}_2$ (light blue in Figure \ref{fig: topology}b)), its interface is oriented along $\hat{\epsilon}_1$, i.e. at $\pi$/3 angle, when it neighbours with a $\hat{\epsilon}_3$ domain (yellow), whereas it is oriented along $\hat{\epsilon}_3$, i.e. at 2$\pi$/3 angle when it neighbours with a $\hat{\epsilon}_1$ domain (red), in agreement with the nanotexture reported in Fig. \ref{fig: topology}. The Saint-Venant condition corresponds to a fixed phase jump $\delta \phi$=2$\pi$/3 of $\overrightarrow{\epsilon}(\mathbf{r})$ across any interface between two monoclinic domains. We note that this condition is easily satisfied throughout the domains of the V$_2$O$_3$ sample, except for the vertex of the V-shaped structures formed by the merging of two domains with $\hat{\epsilon}_2$ order parameter direction (blue) and boundaries oriented along the $\hat{\epsilon}_1$ and $\hat{\epsilon}_3$ directions. If we consider a circuit $\Gamma_1$ across the boundary between two striped domains, the total phase shift is given by $\delta \phi$=+2$\pi$/3-2$\pi$/3=0 thus respecting the curl-free condition. In contrast, the topology of the V-shaped structure is such that, if we move around the internal apex ($\Gamma_2$), the total phase-shift is constrained to $\delta \phi$=+2$\pi$/3+2$\pi$/3=4$\pi$/3 thus breaking the curl-free condition. The direct consequence is that the vertex of the V shaped domains acts as a topological defect with fractional Hopf index (see Supplementary Information Section S6). These topological defects are inherently characterized by the strong frustration of the local value of the order parameter $\overrightarrow{\epsilon} (\mathbf{r})$ and local fluctuations on spatial and temporal scales that cannot be captured by the present experiment.
We further note that the formation of this kind of topological defect is a direct and unavoidable consequence of the quasi-1D confined geometry of the system. Whereas the component of the order parameter parallel to the electrodes ($\epsilon_{||}$, see Fig. \ref{fig: topology}b) can be compensated outside the gap, the perpendicular component ($\epsilon_{\perp}$) should be minimized to avoid excessive strain energy accumulation within the gap region. Thus, considering the directions of $\overrightarrow{\epsilon} (\mathbf{r})$ at the boundaries between different monoclinic domains (see Fig. \ref{fig: topology}b), the formation of V-shaped domains is the only configuration that fulfils the requirement $\epsilon_{\perp}$=0.

\begin{figure*}[t]
\includegraphics[width = 13 cm]{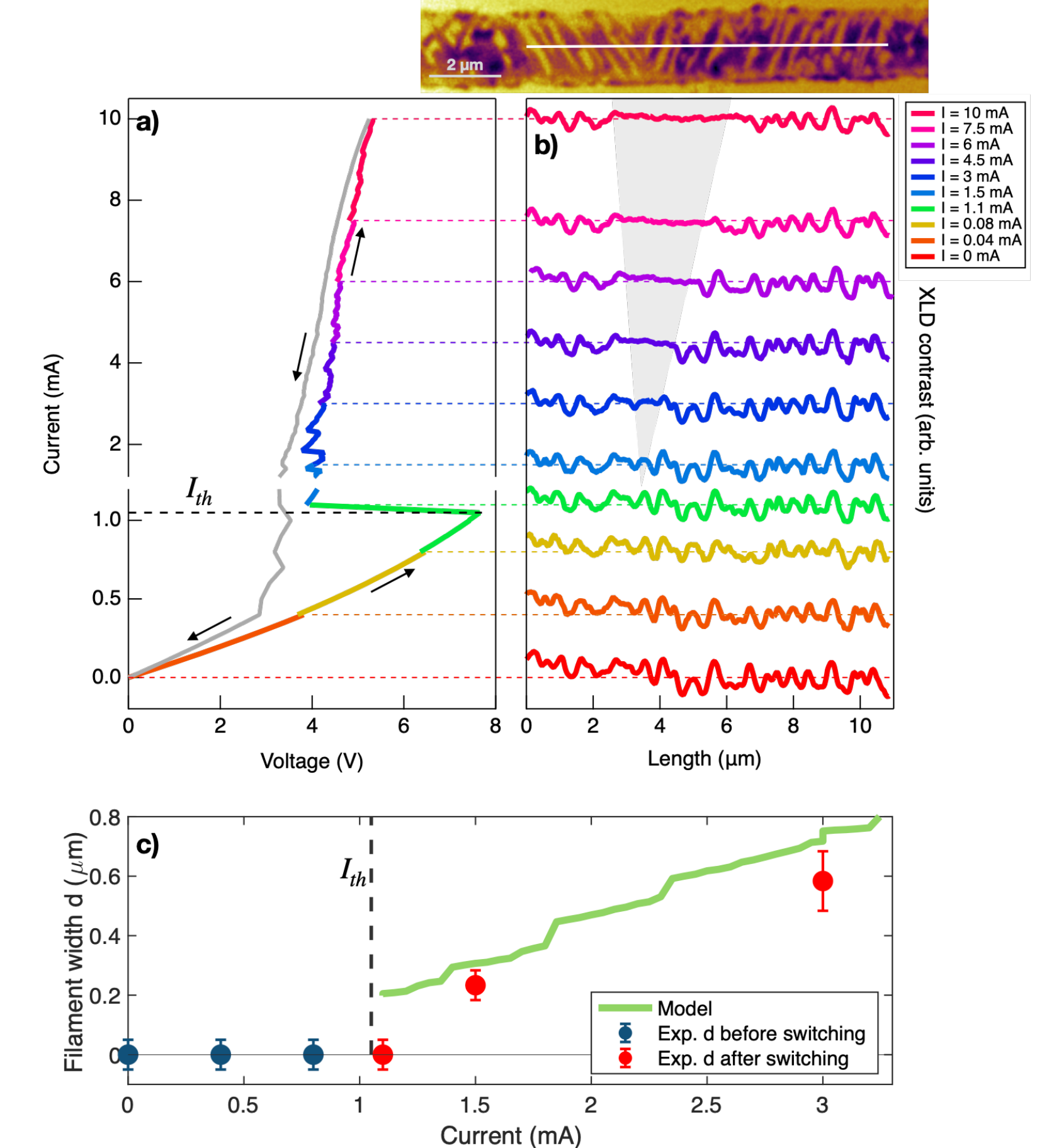}
\caption{ a) I-V curve measured along with the collection of the PEEM images. The drop in voltage measured at I = 1.05 mA signals the first resistive switching. b) Line profiles of the PEEM images in Fig. 2. The grey shaded area indicates the region where the rhombohedral metallic channel is formed. c) Width $d$ of the metallic filament as a function of current $I$ in proximity of $I_{th}$. The blue and red markers represent the values of $d$ obtained experimentally from PEEM images. The green solid line is the estimate of the width $d$ according to the two parallel resistors model, which predicts a jump to a 200 $\textmu$m wide filament at $I_{th}$ (see Supplementary Information Section S5).}
\label{fig: profiles}
\end{figure*}

The suppression of the symmetry-breaking order parameter, $\overrightarrow{\epsilon} (\mathbf{r})$, at topological defects has far-reaching implications related to the nature of the resistive switching process. The electronic IMT can be described by a scalar order parameter $\eta(\mathbf{r})$ \cite{ronchi2022nanoscale}, which depends on the position $\mathbf{r}$ and is such that $\eta=-1$ in the metallic state and $\eta=+1$ in the insulating state. The coupling between the electronic and structural transitions can be described by the energy functional \cite{ronchi2022nanoscale}:
\begin{equation}
F\left[\epsilon,\eta\right] \propto \!\int \!d\mathbf{r}\,\Big\{ 
\big(\eta^2(\mathbf{r})-1\big)^2
- g\big(\epsilon^2(\mathbf{r})-\epsilon^2_{t}(V)\big)\,\eta(\mathbf{r})
\Big\}\,,
\label{E-eta}
\end{equation}
where $g$ is the coupling between the electronic order parameter and the strain and $\epsilon_{t}(V)$ is a threshold parameter that controls the first-order IMT and can depend on the applied voltage $V$. When $\epsilon^2(\mathbf{r})>\epsilon^2_{t}(V)$, the insulating phase with $\eta=+1$ is locally favoured, whereas for strain smaller than the threshold value, i.e. $\epsilon^2(\mathbf{r})<\epsilon^2_{t}(V)$, the metallic solution is stabilized. $\epsilon^2_{t}(V)$ thus represents the threshold above which the insulating monoclinic state ($\eta=+1$, $\epsilon \neq 0$) becomes stable. The description of the electric-field induced transition is based on the observation \cite{mazza2016field} that the electric field directly couples to the electronic bandstructure of a Mott insulator and makes the metallic phase more stable. The transition can thus be described assuming that $\epsilon^2_{t}(V)$ increases as the voltage $V$ is increased. The energy difference between the insulating and metallic phase can be expressed as $\Delta(\mathbf{r},V)= F[-1]-F[+1] \simeq g\left[\epsilon^2(\mathbf{r})-\epsilon^2_{t}(V)\right]$. If we start from the insulating phase with $\epsilon^2(\mathbf{r})>\epsilon^2_{t}(V=0)$, the IMT takes place when $V$ is increased up to the point that  $\Delta(\mathbf{r},V)$=0. A topological defect, which locally suppresses $\epsilon^2(\mathbf{r})$, thus naturally acts as a seed with lower threshold as compared to the rest of the system.

Intriguingly, we also note from Eq. \ref{E-eta} that the IMT can take place at a non-zero value of $\epsilon^2(\mathbf{r})$, which makes possible the formation of a non-thermal metallic state ($\eta=-1$) with finite monoclinic distortion ($\epsilon^2(\mathbf{r})\lesssim \epsilon^2_{t}(V)$), as already observed in non-equilibrium experiments \cite{ronchi2022nanoscale,ronchi2021light}.
The nature of the early-stage switching process can be inferred by a direct comparison between the electrical state of the device and the melting of the monoclinic domains.
The $I-V$ curve of the device, as measured in-situ during the PEEM experiment, is plotted in Fig. \ref{fig: profiles}a). PEEM images are recorded at specific values of $I$. The $I-V$ plot shows that the first resistive switching event occurs at the threshold current $I_{th}$ = 1.05 mA. In Figure \ref{fig: profiles}b) we report a linecut of the PEEM image acquired at specific values of $I$; the image profile is taken along a line crossing the monoclinic domains in the middle of the device gap (see white solid line in Fig. \ref{fig: profiles}, top panel). For large currents running through the device, the line profile in Fig. \ref{fig: profiles}b) displays a flat region, which indicates the melting of the monoclinic nanodomains due to the formation of the rhombohedral metallic channel. As highlighted by the grey area in Fig. \ref{fig: profiles}b), the width $d$ of the metallic filament increases with the current, from $d = 0.23 \pm 0.05$ \textmu m at $I$ = 1.5 mA to $d = 3.7 \pm 0.2$ \textmu m at $I$ = 10 mA.

Modelling the device as a circuit with two parallel resistors (see Supplementary Information Section S5) allows us to estimate the expected width $d$ of the rhombohedral filament corresponding to the observed voltage drop. For large currents running through the device, the experimental values of $d$ match well with what is expected to be the case  when  the metallic channel forming in the gap has the resistivity of the high-temperature rhombohedral phase, as shown in Fig. \ref{fig: profiles}. However, in correspondence of the first resistive switching event at $I_{th}$=1.05 mA, this model already predicts the sudden formation of a $\sim 200$ nm wide metallic rhombohedral filament, which is not visible from PEEM measurements (see Fig. \ref{fig: profiles}c and Supplementary Information Fig. S5), despite being well above the experimental resolution of the microscope. 
To explain this discrepancy, one might suspect that a rhombohedral metallic filament forms below the surface of the \ch{V_2O_3} film, where it is not detected by PEEM which is mainly a surface technique, sensitive to the first few nanometers. In fact, two arguments act against this possibility:
i) the presence of the \ch{Cr_2O_3} buffer layer allows to reduce the substrate-film lattice mismatch from $4.2\%$ to $0.1\%$, thus removing almost entirely the residual epitaxial strain in the V$_2$O$_3$ film \cite{Dillemans2014}, which is known to suppress the monoclinic phase and favour interfacial metallicity \cite{Dillemans2014,pofelski2023domain}. In contrast to highly-strained films, in which the metal to insulator resistivity jump is strongly suppressed \cite{pofelski2023domain}, our films display the 5-order of magnitude resistivity change typical of the unstrained metal-to-insulator transition (see Fig. S1); ii) the curl-free conditions force the interface between monoclinic and rhombohedral metallic regions to be oriented perpendicularly to the order parameter of the monoclinic domain. The formation of a below-surface metallic layer would lead to a sharp ($\ll$ 20 nm) monoclinic-rhombohedral interface parallel to $\overrightarrow{\epsilon}$, thus leading to a dramatic increase of the strain energy of the system.
Our results are compatible with a complex scenario in which the topology-driven resistive switching likely occurs first via the sudden transformation of a single 200 nm wide insulating monoclinic domain into a metallic channel with a non-thermal monoclinic lattice structure. At a second stage, the Joule heating leads to the thermally driven monoclinic-to-rhombohedral structural transition and the formation of rhombohedral metallic channels perpendicular to both the metallic electrodes and the $\hat{\epsilon}_2$ order parameter direction, as observed in Fig. \ref{fig: peem}. 

The X-ray-based nanoimaging of a Mott device under operating conditions allowed us to simultaneously capture the formation of nanoscale conductive paths and the topology of the underlying symmetry-broken nanotexture. The present results expand our knowledge of the resistive switching process in Mott materials by demonstrating the leading role of inherent topological defects in the firing process. The combination of the methodologies here introduced with nanoscale strain engineering approaches unlocks the gate to new opportunities to manipulate topological defects and control the electronic switching dynamics in real devices, such as Mott-transition-based RRAM \cite{SAWA200828,WANG201963}, Mott memristor \cite{Pickett2013,Yoshida2015,Kumar2017} and artificial neurons \cite{Tesler2018,Zhang2020}. The concept of topology-driven resistive switching will be key to assessing the possible non-thermal nature of the early stage electronic phase \cite{ronchi2022nanoscale} as well as the microscopic origin of memory and non-volatile effects recently observed in Mott devices \cite{delValle2019}. We finally note that the intimate relation between topological defects and electronic phase transitions is a general concept, potentially extendable to any system that undergoes a first-order phase transition accompanied by some form of symmetry breaking, as described by the energy functional (\ref{E-eta}). Relevant examples embrace transition-metal oxides \cite{Imada1998,Tokura2017}, such as vanadates, nickelates and manganites, and layered materials, such as 1$T$-TaS$_2$ \cite{Vaskivskyi2015,Hollander2015,Lee2019,Gao2022}, in which the insulator-to-metal transition is accompanied by charge-, lattice- and orbital-ordered states with reduced symmetry. Other interesting platforms are cuprate superconductors \cite{Keimer2015} and kagome metals \cite{Asaba2024} in which light- or magnetic-induced discontinuous electronic transitions coexist with charge-order. We thus provide a new framework for non-equilibrium electronic phase transitions and optical control of hidden states \cite{Stojchevska2014,Zhang2016,Gao2022} in a broad class of quantum materials, in which the early-stage dynamics can be reversibly controlled by light or voltage, exploiting topological defects of the underlying symmetry-breaking order parameter \cite{Wandel2022,Gao2022,Cheng2024}.
\\
\\
We thank Diamond Lights Source for the provision of beamtime under proposal numbers MM-27218, MM-31711 and MM-34455. 
We thank Manuel R. Osorio and Fernando J. Urbanos for the fabrication of sample electrodes at the Centre for Micro and Nanofabrication of IMDEA Nanociencia. 
A.M., S.M. and C.G. acknowledge financial support from MIUR through the PRIN 2015 (Prot. 2015C5SEJJ001) and PRIN 2017 (Prot. 20172H2SC4\_005) programs and from the European Uninion - Next Generation EU through the MUR-PRIN2022 (Prot. 20228YCYY7) program. 
Funded by the European Union - Next Generation EU
C.G. acknowledges support from Università Cattolica del Sacro Cuore through D.1, D.2.2 and D.3.1 grants. S.M. acknowledges partial financial support through the grant ”Finanziamenti ponte per bandi esterni” from Università Cattolica del Sacro Cuore.
I.F.C. and M.M. acknowledge support from the “Severo Ochoa” Programme for Centres of Excellence in R\&D (CEX2020-001039-S) and the Spanish AEI-MCIN PID2021-122980OB-C52 (ECoSOx-ECLIPSE).  I.F.C holds a FPI fellowship from the Spanish AEI-MCIN (PRE2020-092625). 
W.-F.H., S.M., J.W.S. and J.-P.L. acknowledge financial support by the KU Leuven Research Funds Project No. C14/21/083, iBOF/21/084, KAC24/18/056 and C14/17/080, as well as the FWO AKUL/13/19 and AKUL/19/023, and the Research Funds of the INTERREG-E-TEST Project (EMR113) and INTERREG-VL-VL-PATHFINDER Project (0559).

\bibliography{Refs}
\end{document}


\title{Supplementary Information for \\ Topological resistive switching in a Mott insulator}

\author{Alessandra Milloch}
\email[]{alessandra.milloch@unicatt.it}
\affiliation{Department of Mathematics and Physics, Università Cattolica del Sacro Cuore, Brescia I-25133, Italy}
\affiliation{ILAMP (Interdisciplinary Laboratories for Advanced
Materials Physics), Università Cattolica del Sacro Cuore, Brescia I-25133, Italy}
\affiliation{Department of Physics and Astronomy, KU Leuven, B-3001 Leuven, Belgium}

\author{Ignacio Figueruelo-Campanero}
\email[]{ignacio.figueruelo@imdea.org}
\affiliation{IMDEA Nanociencia, Cantoblanco, 28049 Madrid, Spain}
\affiliation{Facultad Ciencias Físicas, Universidad Complutense, 28040 Madrid, Spain}

\author{Wei-Fan Hsu}
\affiliation{Department of Physics and Astronomy, KU Leuven, B-3001 Leuven, Belgium}

\author{Selene Mor}
\affiliation{Department of Mathematics and Physics, Università Cattolica del Sacro Cuore, Brescia I-25133, Italy}
\affiliation{ILAMP (Interdisciplinary Laboratories for Advanced
Materials Physics), Università Cattolica del Sacro Cuore, Brescia I-25133, Italy}

\author{Simon Mellaerts}
\affiliation{Department of Physics and Astronomy, KU Leuven, B-3001 Leuven, Belgium}

\author{Francesco Maccherozzi}
\affiliation{Diamond Light Source, Didcot, Oxfordshire OX11 0DE, UK}

\author{Larissa {Ishibe Veiga}}
\affiliation{Diamond Light Source, Didcot, Oxfordshire OX11 0DE, UK}

\author{Sarnjeet S. Dhesi}
\affiliation{Diamond Light Source, Didcot, Oxfordshire OX11 0DE, UK}

\author{Mauro Spera}
\affiliation{Department of Mathematics and Physics, Università Cattolica del Sacro Cuore, Brescia I-25133, Italy}

\author{Jin Won Seo}
\affiliation{Department of Materials Engineering, KU Leuven, 3001 Leuven, Belgium}

\author{Jean-Pierre Locquet}
\affiliation{Department of Physics and Astronomy, KU Leuven, B-3001 Leuven, Belgium}

\author{Michele Fabrizio}
\affiliation{Scuola Internazionale Superiore di Studi Avanzati (SISSA), Via Bonomea 265, 34136 Trieste, Italy}

\author{Mariela Menghini}
\affiliation{IMDEA Nanociencia, Cantoblanco, 28049 Madrid, Spain}

\author{Claudio Giannetti}
\email[]{claudio.giannetti@unicatt.it}
\affiliation{Department of Mathematics and Physics, Università Cattolica del Sacro Cuore, Brescia I-25133, Italy}
\affiliation{ILAMP (Interdisciplinary Laboratories for Advanced
Materials Physics), Università Cattolica del Sacro Cuore, Brescia I-25133, Italy}
\affiliation{CNR-INO (National Institute of Optics), via Branze 45, 25123 Brescia, Italy}

\maketitle

\section{\ch{V_2O_3} film resistivity}
\begin{figure}[h]
\includegraphics[width = 7 cm]{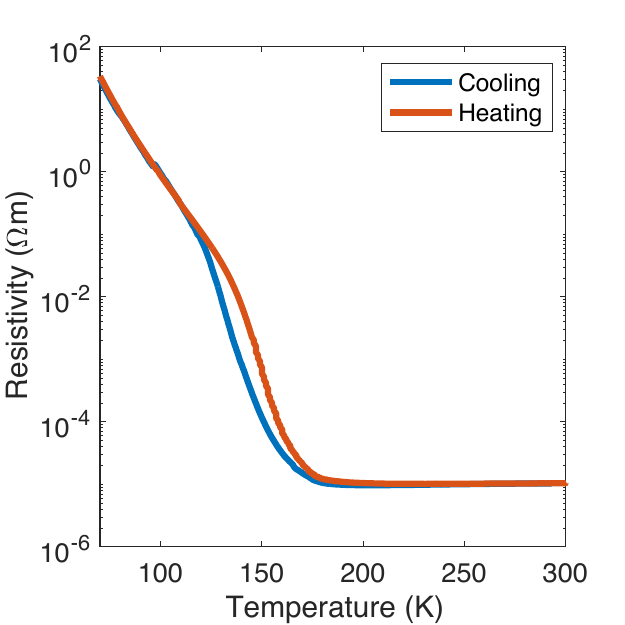}
\caption{Temperature dependent resistivity of the 20 nm \ch{V_2O_3} film employed for PEEM measurements.}
\label{fig: resistivity}
\end{figure}

\section{Device fabrication}
The \ch{V_2O_3} thin film is patterned using standard optical lithography using AZ1505 resist. After developing the exposed areas, we deposit 5 nm of \ch{Ti} + 60 nm of \ch{Au} in an electron beam evaporator with a base pressure in the range of $10^{-9}$ mbar. Finally, the sample is lifted off in acetone and cleaned with IPA and \ch{N_2} gas. 

\newpage
\section{X-ray absorption spectroscopy}

\begin{figure}[!h]
\includegraphics[width = 12 cm]{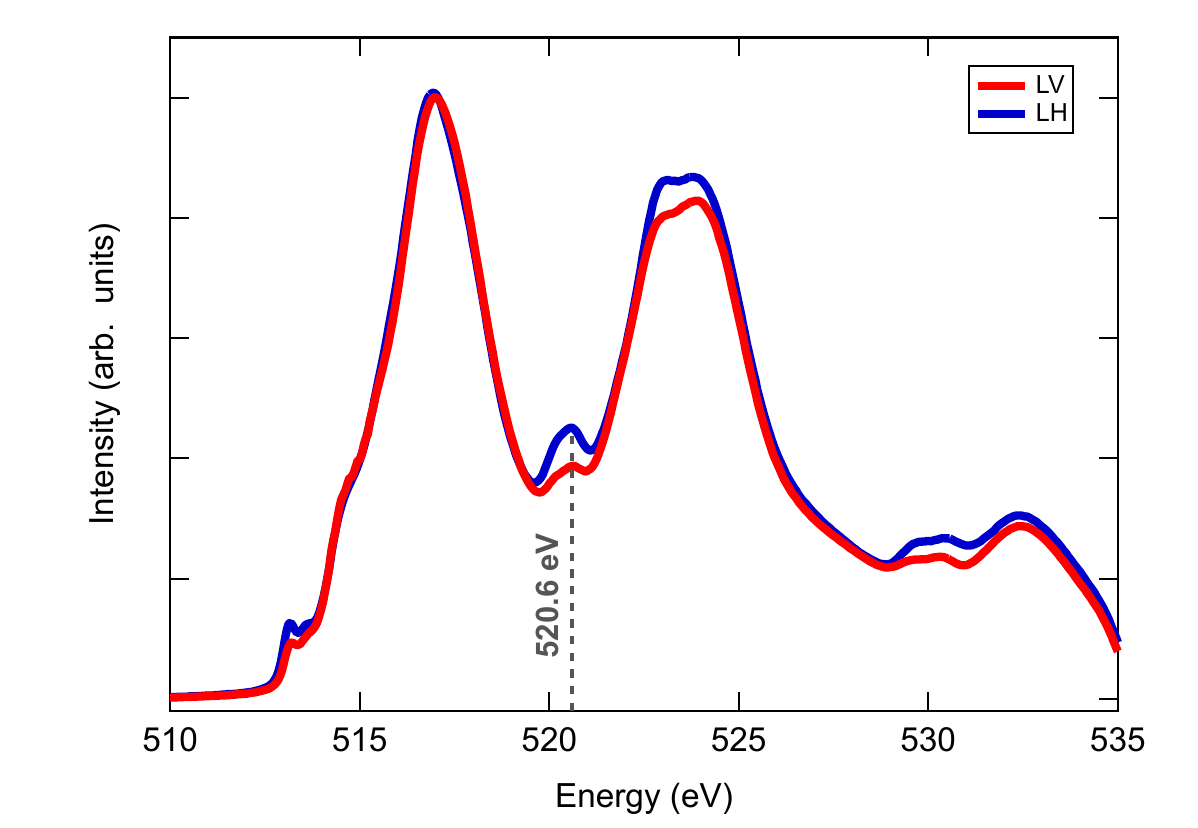}
\caption{X-ray absorption spectra at the vanadium L-edge for linear vertical (LV) and linear horizontal (LH) polarizations. PEEM imaging is performed by exploiting the linear dichroism at photon energy 520.6 eV.}
\label{fig: resistivity}
\end{figure}
\FloatBarrier

\section{PEEM imaging during resistive switching}

\begin{figure}[h]
\includegraphics[width = 16 cm]{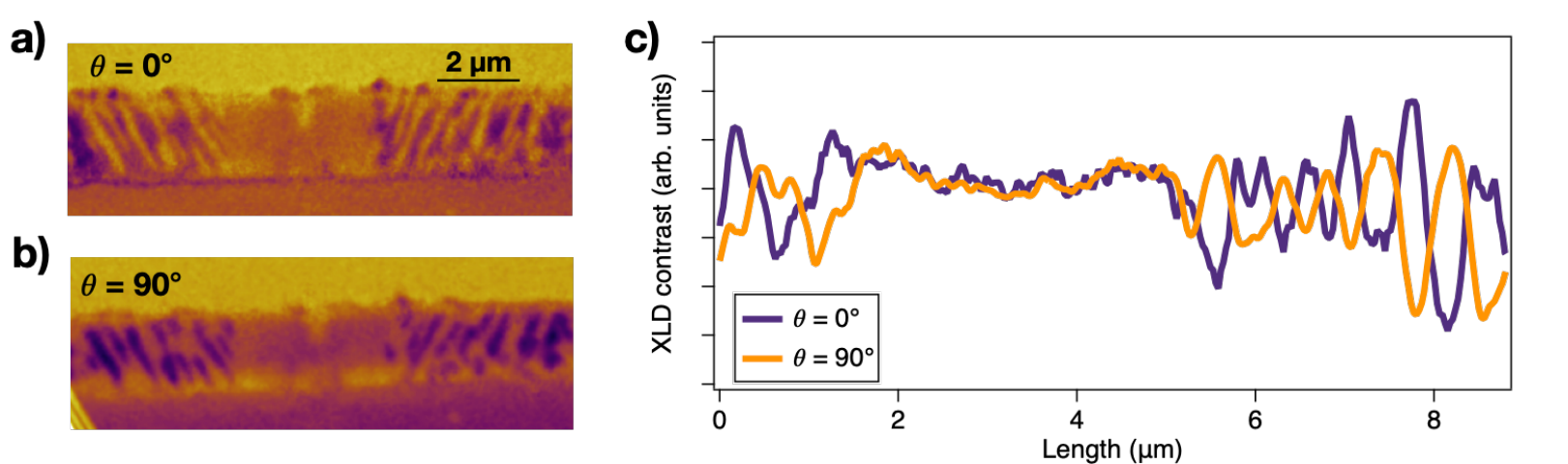}
\caption{a, b) PEEM images collected for two orthogonal directions of the X-rays polarization with respect to the crystalline axes of the \ch{V_2O_3} film. Upon rotation, the XLD signal is unchanged in the untextured corundum region in the middle of the image while it changes sign in the monoclinic region, with dark domains in a) appearing as light domains in b) and vice versa. c) Line profile of the PEEM images in a) and b).}
\label{fig: rotation}
\end{figure}

\begin{figure}[h]
\includegraphics[width = 16 cm]{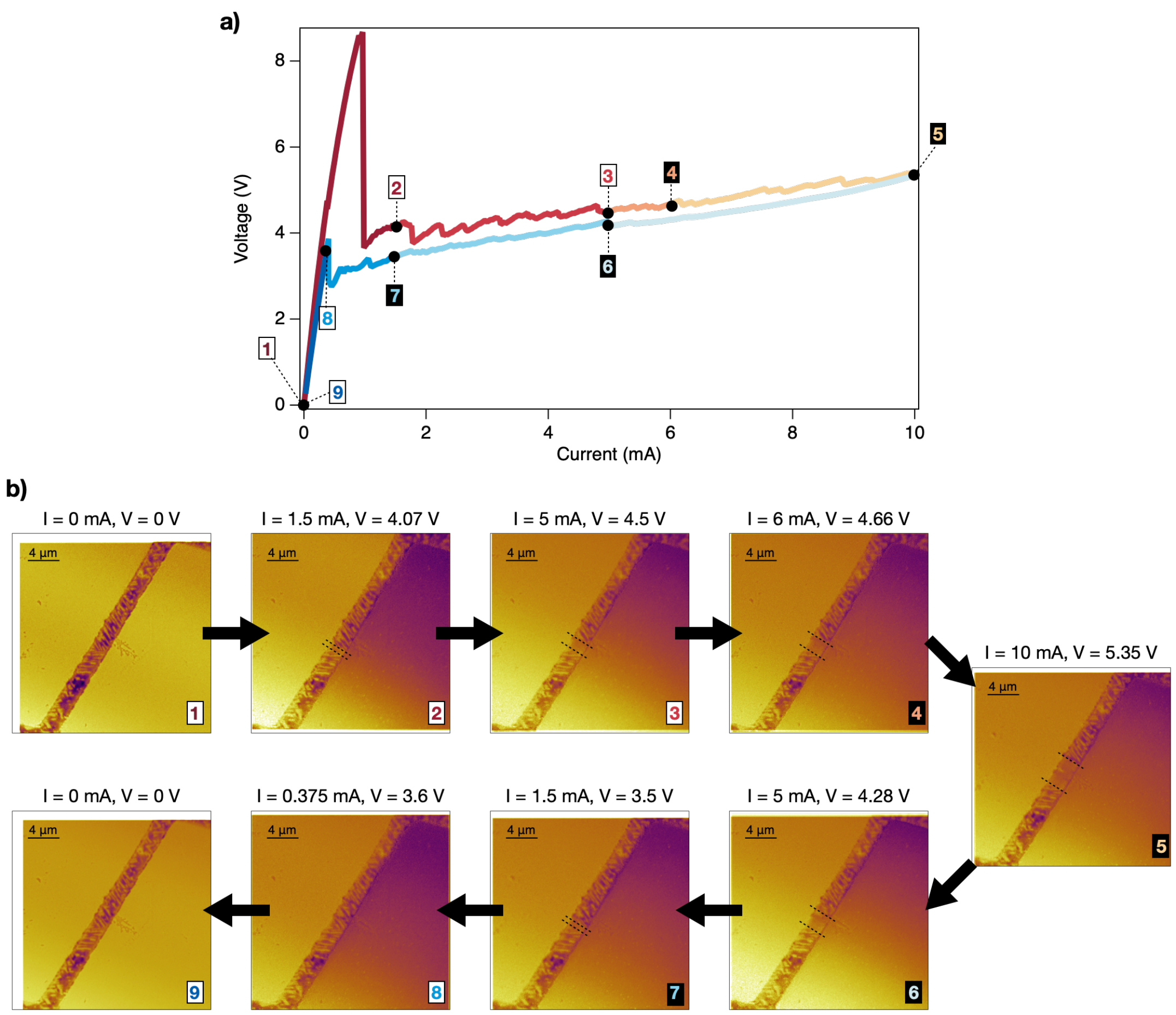}
\caption{a) $I-V$ curve obtained from a current sweep from 0 to 10 mA (red lines) and back from 10 to 0 mA (blue lines). The colored numbers in the plot indicate the point in the $I-V$ curve where PEEM imaging is performed while keeping constant the current $I$ running though the device. b) PEEM images collected on the device gap while an electric current is applied. The 25 \textmu m field of view employed here allows to image the full device length. The corundum filament formed after resistive switching is detected for currents $I >$ 1.5 mA and is highlighted by black dashed lines. When the current is decreased, the corundum filament disappears at $I = 0.375$ mA (panel 8), i.e. just outside of the hysteretic region of the I-V curve.}
\label{fig: fov 25 um}
\end{figure}

\clearpage
\section{Metallic channel width estimate}

The \ch{V_2O_3} device can be modelled as a circuit with two parallel resistors. The corundum channel is described as a wire of resistance $R_M = \rho_M w / dt$, where $\rho_M$ is the \ch{V_2O_3} resistivity of the high-temperature phase, $w$ is the separation between the two electrodes, $d$ is the width of the metallic filament formed after resistive switching and $t$ is the vanadium oxide film thickness. The region of the device gap that remains in the monoclinic phase is instead assumed to have resistivity $\rho_I = R_I (l - d) t / w$, with $R_I$ resistance measured from a two-point measurement at $T$ = 120 K. 

\begin{figure*}[p]
\includegraphics[width = 15 cm]{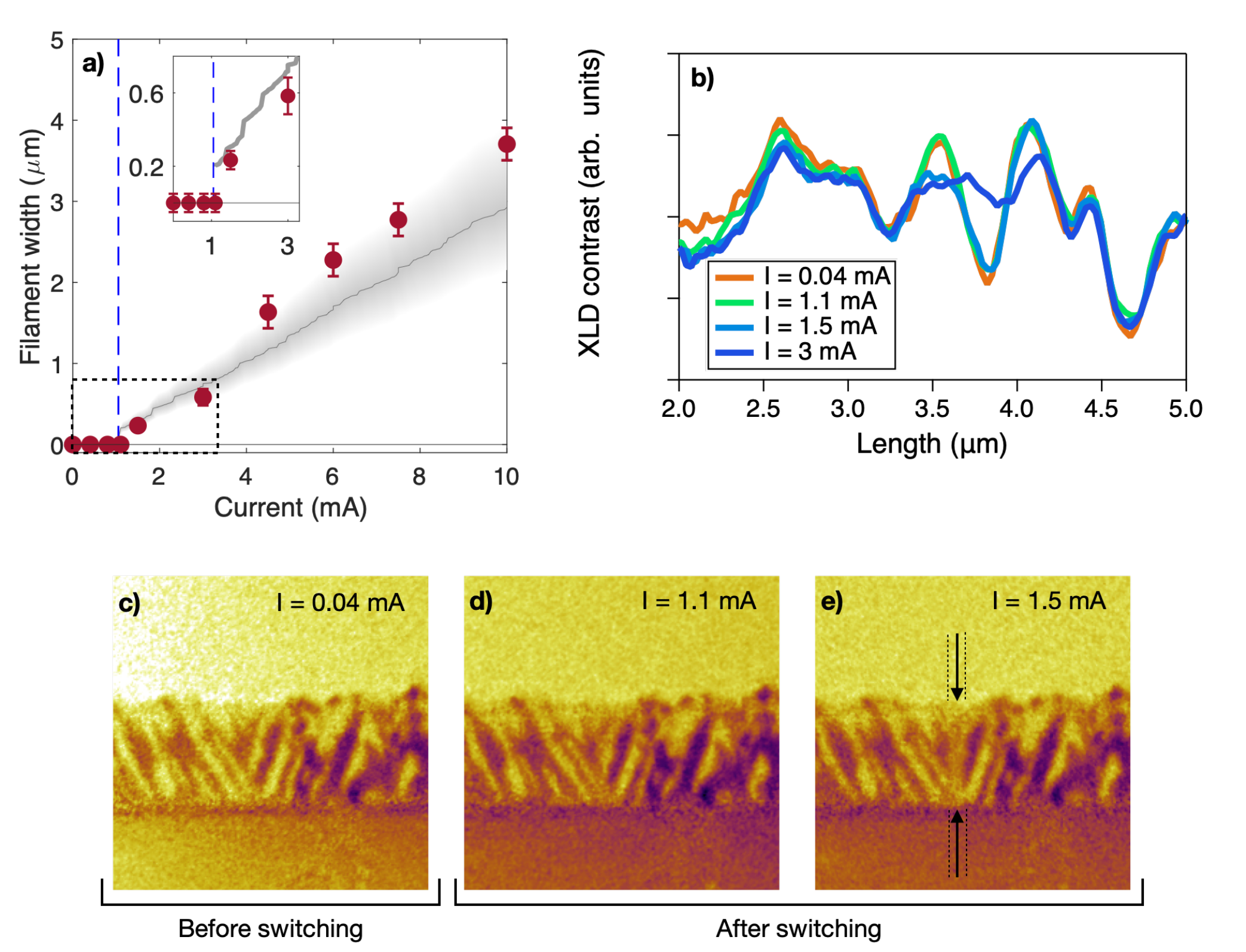}
\caption{Details of the filament formation across the resistive switching threshold current.  a) Width of metallic filament (red markers) as a function of the current $I$. The blue dashed line indicated the threshold current for resistive switching 1.05 mA. The grey line is the estimated width of a metallic filament obtained from a two parallel resistors model. The inset represents a zoom-in of the area highlighted by the black dashed rectangle. b) Line profiles of the PEEM images where the metallic filament forming at position $\sim$3.5 \textmu m is detected by the flattening of XLD contrast. No variation is observed right after resistive switching, but only starting from $I$ = 1.5 mA. c-e) PEEM images collected with applied currents lower (c) and higher (d-e) than 1.05 mA where resistive switching happens. No 200 nm wide corundum filament is observed right after the voltage drop (d).}
\label{fig: non thermal switching}
\end{figure*}

The expected metallic filament width is estimated considering the instantaneous equivalent resistance $R_{eq}$ - obtained from the $I-V$ curve in Fig. 4a and taking into account the estimated contact resistance of 200 $\Omega$ - and $\rho_M = 0.001 ~\Omega$ cm, which is the value for the high-temperature resistivity of \ch{V_2O_3} thin films (see fig. S1). The values obtained from the two resistors model are plotted in Figure \ref{fig: non thermal switching}a (grey solid line) as a function of current $I$. The shaded area represents the range of the values estimated for $d$ when $\rho_M$, $t$ and $l$ are varied by 10$\%$ to account for experimental uncertainty. For $I > I_{threshold}$, this simple model accounts for the filament width measured experimentally (red markers in Fig. \ref{fig: non thermal switching}a). In proximity of the resistive switching, however, there is no agreement between the values of $d$ estimated from the model and the experiment. In particular, as highlighted in the plot in the inset, at $I_{threshold}$ the two resistors model predicts the formation of a $200 \pm 50$ nm wide filament, which, despite being well above the experimental resolution of the microscope, is not detected from PEEM measurements.
 Figure \ref{fig: non thermal switching}c-e display a detail of the PEEM images in the region of the filament formation. In the first image acquired at $I>I_{threshold}$ (fig. \ref{fig: non thermal switching} d), namely at $I$ = 1.1 mA which corresponds to the point in the $I-V$ curve at the base of the voltage drop, the nanodomains arrangement appears unchanged as compared to the configuration observed before switching (fig. \ref{fig: non thermal switching}c). This is better highlighted from the comparison of the PEEM line profiles reported in fig. \ref{fig: non thermal switching}b. The XLD signal modulations obtained at $I$ = 0.04 mA and $I$ = 1.1 mA overlap very well, as opposed to the curves for $I\geq$1.5 mA that deviate from the before-switching ones in the 3.5 \textmu m region where the corundum filament is formed.

\clearpage
\section{Orbifolds}

The $4\pi/3$ phase shift of $\vec{\epsilon}$ (or, rather, a suitable regularization thereof) around the circuit $\Gamma_2$ with ensuing {\it fractional} Hopf index (at the vertex $V$ of the v-shaped domain) $\iota_V({\vec{\epsilon}}) = (4\pi/3) : 2\pi = 2/3$ suggests the presence of a $2\pi/3$-{\it conical singularity} at $V$:
a neighbourhood of $V$ is a {\it cone} obtained by identifying the sides of a circular sector  with angle $2\pi/3$, or, equivalently, as the orbit space $D/{\mathbb Z}_3$, $D$ being a disc centred at $V$ and the ${\mathbb Z}_3$-action is generated by the $2\pi/3$-rotation around $V$, giving
rise to an {\it orbifold} ${\mathcal O}$ (see e.g. \cite{satake1956generalization,satake1957gauss,seaton2008two,caramello2019introduction}). Specifically, this goes as follows. The fundamental group $ \pi_1(D \setminus \{ V\} \cong {\mathbb Z})$ can be represented via the holonomy of a flat $U(1)$-connection (gauge field)
${\tilde{\epsilon}} = \hat{\epsilon} \, d\phi$,  $\hat{\epsilon} = 2/3$, gauge equivalent to ${\epsilon} = \epsilon_1 dx + \epsilon_2 dy$ (where $\vec{\epsilon} = (\epsilon_1, \epsilon_2)$), namely
$$
U(\gamma) := e^{i\,\int_{\gamma} \epsilon} = e^{i\,\int_{\gamma} \tilde{\epsilon}} 
$$
yielding, for $\gamma = \Gamma_2$, 

$$
U(\Gamma_2) = e^{2\pi i\cdot \frac{2}{3}} = e^{ \frac{4}{3} \pi i} 
$$
reproducing the overall phase shift of $ \vec{\epsilon}$. The representation $U$ then descends to a representation the fundamental group of the cone (with $V$ removed) ($\cong {\mathbb Z}_3$) since $U(\Gamma_2^3) = U(\Gamma_2)^3 = 1$.\par
Globally, one may envisage the following situation: if we
assume ${\mathcal O}$ compact and topologically a sphere $S$, then ${\mathcal O}$ could be taken as  the quotient $S/{\mathbb Z}_3$ (with the ${\mathbb Z}_3$-action given by rotations around a fixed axis): this is an example of {\it good} orbifold.
The {\it orbifold Euler characteristic} $\chi^{\rm orb} ( {\mathcal O})$ then equals $\chi(S)/3 = 2/3$. Indeed, recall that
if $|{\mathcal O}|$ denotes the topological space underlying ${\mathcal O}$ and the latter possesses a finite number of $2\pi/p_j$-conical singularities $o_j$, the following general formula 
holds:
$$
\chi^{\rm orb} ( {\mathcal O}) = \chi(|{\mathcal O}|) - \sum_j (1 - \frac{1}{p_j})
$$
together  with the {\it orbifold Poincar\'e - Hopf theorem}
$$
\chi^{\rm orb} ( {\mathcal O}) = \sum_j \iota^{\rm orb}_{o_j} (X)
$$
with $X$ any smooth vector field on ${\mathcal O}$ and
$$
\iota^{\rm orb}_{o_j} (X) = \iota_{o_j} (X) / p_j.
$$
where $\iota_{o}(\cdot)$ is a standard Hopf index (total phase shift when traversing a simple circuit around $o$ divided by $2\pi$ = number of turns of a vector field around $o$). We are then led to view $\vec{\epsilon}$ as a vector field with a zero at $V$, with index $2/3$ and no other singularity.
 On the other hand, a single {\it teardrop} singularity $(\chi^{\rm orb} ( {\mathcal O}) = 4/3$, a {\it bad} orbifold) is not allowed unless $ \vec{\epsilon}$ also vanishes at another point, with index $2/3$ thereat.\par

\bibliography{Refs}